# Evolution of the double neutron star merging rate and the cosmological origin of gamma-ray burst sources

V.M.Lipunov[1,2], K.A.Postnov[1,2], M.E.Prokhorov [2] and I.E.Panchenko[1]

Faculty of Physics, Moscow University, 117234 Moscow, Russia

and

Sternberg Astronomical Institute, Moscow University, 119899 Moscow, Russia

H.E. Jorgensen[3]

Astronomical Observatory, Ostervoldgade 3, DK-1350 Copenhagen, Denmark

[1]e-mail: lipunov@sai.msu.su

[3]e-mail: henning@astro.ku.dk

## ABSTRACT

Evolution of the coalescence rate of double neutron stars (NS) and neutron star – black hole (BH) binaries are computed for model galaxies with different star formation rates. Assuming gamma-ray bursts (GRB) to originate from NS+NS or NS+BH merging in distant galaxies, theoretical $\log N$–$\log S$ distributions for gamma-ray bursts (GRB) are calculated for the first time taking the computed merging rates into account. We use a flat cosmological model ($\Omega = 1$) with different values of the cosmological constant $\Lambda$ and under various assumptions about the star formation history in galaxies. The calculated source evolution predicts a 5-10 times increase of the source statistics at count rates 3-10 times lower than the exising BATSE sensitivity limit. The most important parameter in fitting the 2nd BATSE catalogue is the initial redshift of star formation, which is found to be $z_* = 2 - 5$ depending on a poorly determined average spectral index of GRB.

*Subject headings:* Binaries: close – cosmology: miscellaneous – gamma rays: bursts – stars: neutron

## 1. Introduction

The observed isotropy on the sky and non-uniform spatial distribution of GRB revealed by the BATSE device onboard Compton Gamma-Ray Observatory (e.g. Meegan et al. 1992) lend support to the idea of cosmological origin of GRB (Prilutski & Usov 1975; Usov & Chibisov 1975) for which the best candidates could be merging binary NS+NS or NS+BH at high redshifts ($z \simeq 1-2$), (Blinnikov, et al. 1984; Paczyński 1991, 1992). Narayan, Paczyński & Piran (1992) argued that the rate of double NS coalescence as a result of orbital shrinking induced by gravitational waves are far from being sufficient to explain the observed properties of cosmic GRB and their rate ($\sim 0.8$ bursts per day). Wickramasinghe et al (1993) showed the consistency of standard cosmology and the BATSE GRB $\log N$–$\log S$ distribution.

The cosmological models for GRB have not yet been proven; moreover, they come across severe problems (such as no-host-galaxy limits, baryon contamination degradation of the high energy photons, efficiency problems at getting the NS binding energy out of the BH into gamma-rays etc.), which we will not address here.

However, the cosmological models involving binary NS or NS+BH coalescences must have clear observational consequences in showing the effects of binary coalescence rate evolution on the observed $\log N$–$\log S$ distribution and $\langle V/V_{max} \rangle$. Attempts to take the intrinsic evolution of the sources into account have been carried out in a number of papers (see e.g. Piran 1992; Yi 1994; Cohen & Piran 1995), but all of them used simple *ad hoc* assumptions for the sources evolution.

Both the parameters of the cosmological model and source evolution are known to influence the shape of the integral statistical distributions of sources (e.g. Weinberg 1972), and it has until now been very difficult or even impossible to separate these effects from each other. This is similar to the failure of using counts of radio sources to check the cosmological models, as the evolution of number per comoving frame, spectral shape, luminosity etc is very complicated and poorly understood as yet.

In contrast, the evolution of binary systems being based and confirmed by a bulk of astronomical observations at different wavelengths is much better understood. The analysis of the evolutionary scenario of binary systems by Lipunov et al. 1995 showed that a few key parameters largely define the overall binary evolution. These parameters are the spectrum $f(q)$ of the initial binary mass ratio $q = M_2/M_1 \leq 1$ and the efficiency $\alpha_{CE}$ of binary orbital momentum transfer into a common envelope. These parameters can be constrained by comparing numbers of binaries in different stages of evolution in the Galaxy predicted by the scenario with observed numbers (Lipunov et al. 1995).

In the present paper we compute the cumulative statistical distribution, $\log N$–$\log S$ , and $\langle V/V_{max}\rangle$ test for binary NS and NS+BH coalescence taking into account the temporal change of their rates found by Monte-Carlo modelling of the modern scenario of binary star evolution (the "Scenario Machine" method). A comparison of the computed distribution with that of the 2nd BATSE GRB catalogue (Meegan et al. 1994) is also made.

## 2.  Calculation of binary NS coalescence rates

We have calculated the binary coalescence rate with time using the "Scenario Machine" code, which allows us to simulate evolution of large ensembles of binary stars in an artificial galaxy using a Monte Carlo method (see Lipunov 1992, Lipunov et al 1994, 1995 and references therein).

It is reasonable to expect a priori that the evolution of events is strongly time dependent, so the star formation history $\phi(t)$ in a galaxy is another important parameter.

Therefore, by calculating the evolution of events after a $\delta$-function like star formation burst, one gets a Green function $f_e(t)$ for any arbitrary star formation history.

At the one extreme we assume instantaneous star formation at a particular redshift resulting in a stellar system which we call "elliptical" since stars in elliptical galaxies to a good approximation can be assumed to be old and little if any additional star formation is occuring today. Thus the event rates in "ellipticals" can be described by the computed Green function. At the other extreme, constant star formation ($\phi(t) = constant$) would result in "spiral"-like stellar systems. Irregular galaxies have probably also an irregular star formation history, but their contribution to the overall event rate hardly exceeds their fraction among all galaxies, that is $5 - 7\%$, so we neglect them in this context.

The source evolution in a galaxy with given star formation rate $\phi(t)$ is thus calculated as $f_s(z) = \int_{z_*}^{z} f_e(z_* - z')\phi_s(z')dz'$, where $z_*$ is the redshift at the turn-on of star formation, which is the second important parameter of the model. Assuming to the zero approximation $\phi_s = const$, we get $f_s = \int_{0}^{z_*} f_e(z_* - z')dz'/t(z_*)$.

We parametrize the star formation history in the Universe by the fractional part of the luminous baryonic matter entering into elliptical galaxies, $\epsilon = E/(E + S)$, where $E$ and $S$ refer to "elliptical" galaxies (i.e. without additional star formation) and "spiral" galaxies (with a constant star formation rate). In fact, this parameter must be higher than the presently observed fraction of elliptical galaxies, as any galaxy must have had more violent star formation at earlier time. The galaxies are supposed to be formed at the moment $z_*$ with the initial star formation during first 500 million years. The mean number of galaxies each of $10^{11}M_\odot$ was taken $n_G = 0.08$ per cubic megaparsec (this roughly corresponds to a density of baryonic matter in galaxies without hidden mass of $\Omega_v \simeq 0.03$).

We assumed an initial distributions of binary stars similar to those presently observed in our Galaxy (a Salpeter function for mass $M_1$ of the primary component, $f(M_1) \propto M_1^{-2.35}$ and a flat distribution of the initial binary separations $A$: $f(logA) = const$). Based on findings by the "Scenario Machine" analysis of the evolutionary scenario by Lipunov et al. (1995), we take the initial mass ratio distribution in a power-law form $f(q) \propto q^2$ and the efficiency coefficient of angular momentum removal at the common envelope stage $\alpha_{CE} = 1$ (as defined by van den Heuvel 1994).

## 3. Effect of sources evolution on $\log N - \log S$ distribution and $\langle V/V_{max} \rangle$-test.

We conventionally assume a flat Universe so that $\Omega = \Omega_M + \Omega_\Lambda = 1$, with $\Omega_M$ and $\Omega_\Lambda$ being fractional contributions of matter and cosmological constant term (Carroll, Press & Turner 1992). The present value of the Hubble constant is assumed $H_0 = 75$ km/s/Mpc. The GRB were considered as standard candles with a proper luminosity $L$ and have a power-law spectrum $\propto E^{-s}$ with a spectral index $s = 1.5$ (Shaefer et al. 1992). Briefly, the count rate at the detector from a source at a redshift $z$ is

$$C(z) = \frac{L(1+z)^{-s}}{4\pi d_m^2(z)},\tag{1}$$

where $d_m(z)$ is a metric distance (see Carroll et al. 1992). The number of events, $N(>C)$, with an observed count rate exceeding $C$ is thus

$$N(>C) = 4\pi \int_0^{z(C)} d_m^2 \frac{n(z)}{(1+z)} \frac{d(d_m)}{dz} dz,\tag{2}$$

The source evolution $n(z)$ can easily be obtained using the evolutionary Green function: $n(z) = \epsilon f_e(z) + (1-\epsilon)f_s$ and the dependence $t(z)$ for a particular cosmological model.

Fig. 1 shows the "Green functions" for the double compact binary merging rate evolution assuming no collapse anisotropy. A very strong early evolution with time is seen; however, even for an elliptical galaxy the NS+NS merging rate is 1 per $10^4$ yrs for an age of $10^{10}$ yrs. The non-monotonic character of the merging rate evolution is due to different contributions of different type binaries (by initial masses, separations and types of the first mass exchange); as the detailed shape is of less importance for us now, we postpone discussing these interesting features in a separate paper. We should, however, note that a small decrease of the coalescence rate observed at the age of about 4 billion years is statistically significant and is caused by contributions of evolutionary different types of NS+NS binaries (they come from different ranges of the initial mass ratios and semimajor axes). This feature can produce a notable decrease in the $\log N$–$\log S$ curve slope at the corresponding $C_{max}/C_{lim}$. Thus, if $z_* = 2.25$ the drop in sources production rate after 4 billion years from the beginning corresponds (for the flat Universe), somewhat surprisingly, just to the feature visually seen in both 2d and 3d BATSE catalog $\log N$–$\log S$ curves; however, it would be prematurely to take the apparent feature seriously due to its low statistical significance. Another possible way that could lead to arising diverse peculiarities in the $\log N$–$\log S$ distribution induced by the sources evolution is connected with a possible non-monotonic character of the star formation rate in galaxies.

For a $10^{11} M_\odot$ spiral galaxy with constant star formation rate our calculations give a binary NS coalescence rate of the order of one per five thousand years only slightly depending on the initial mass ratio spectrum and assumptions about massive core collapse anisotropy. This rate is close to the most "optimistic" estimates based on evolutionary considerations (Lipunov et al 1987, 1995; Tutukov & Yungelson 1992; van den Heuvel 1994). In our calculations we used a smoothed evolutionary function assuming the star formation to occur during the first 500 million years in elliptical galaxies, and constant in spiral galaxies. In fact, the results proved to be only weakly sensitive to the smoothening over time provided that it is less than 1 billlion years (the width of the wide peak of the Green function; see Fig. 1).

Our theoretical models depend on three unknown parameters, $z_*$, $\Lambda$, $\epsilon$, as well as on $s$.

The calculated $\log N$–$\log S$ were compared with the BATSE data by using Mises-Smirnov $\omega^2$ test (which gives essentially the same results as Kolmogorov-Smirnov test, but uses a more smooth criterion for comparing observed and tested distribution functions; see Bol'shev & Smirnov 1965). Fig. 2a-c show two-dimensional cuts through the parameter space. All three cuts show contour lines for confidence level according to $\omega^2$ test higher than 90%, with the maximum level being at $\approx 95\%$ in all planes.

The calculated $\log N$–$\log S$ distributions are plotted in Fig. 3 for different star formation starting times $z_*$. Other parameters are fixed at the best-fit values $\Omega_\Lambda = 0.75$ and $\epsilon = 0.5$. The most prominent feature of the obtained theoretical $\log N$–$\log S$ is a notable sharp increase at low count rates due to evolutionary effects. In terms of $\langle V/V_{max} \rangle$-test, such a turnup of the $\log N$–$\log S$ curve would lead to a sharp increase of the average $\langle V/V_{max} \rangle$values when weaker sources (i.e. farther sample limits) are taken into account (Fig. 4). The curves turn out to be most sensitive to the parameter $z_*$ and depend only slightly on other parameters. We also note that a flatter GRB spectral index $s = 1$ would favour earlier initial star formation $z_* > 3$ and changes the $\log N$–$\log S$ curves stronger at lower $z_*$.

If the cosmological binary NS coalescences do underly the GRB phenomenon, late epochs of the initial star formation in galaxies ($z_* < 2$) would yield inconsistency with the already existing BATSE data.

## 4. Conclusions

Our calculations show that if GRB indeed are due to binary neutron star coalescence, they can potentially constrain the cosmological parameters as well as the early star formation history. Even for the simplest models we used, late epochs of galaxy formation ($z_* < 2$) do not seem to be consistent with the observed BATSE $\log N$–$\log S$ distribution.

The best-fit model we obtained by fitting to the 411 GRB available from the 2nd BATSE catalogue (Meegan et al. 1994) corresponds to $z_* \simeq 2.2$, $\epsilon = 0.5$ and $\Omega_\Lambda = 0.75$, with a $\approx 95\%$ agreement according to $\omega^2$-test (we also used a Kolmogorov-Smirnov test; both criteria give qualitatively similar results but we present results for the $\omega^2$-test as it gives a more smooth likelihood function). According to the best-fit, the most distant BATSE GRB come from $z_{max} \simeq 1.5$ (compared to Cohen & Piran 1995 who obtained $z_{max} = 2$; however, we made no data selection). The results are also sensitive to the accepted mean spectral index of the GRB ($s = 1.5$ in our case) and favours earlier star formation $z_* > 3$ for flatter spectra ($s = 1$). We also note that the earlier epochs of the primordial star formation are not favored by other cosmological grounds (see calculations by Cen et al. 1994).

One may also wonder how the assumption about the total $\Omega$ change the results. Obviously, we can fit the observations for a wide range of $\Omega$ by varying other parameters ($z_*$, $s$, $\epsilon$ etc.). The dependence of $\log N$–$\log S$ curves on $\Omega$ was found to be rather small (Yi 1994), and our main conclusion still holds – the $\log N$–$\log S$ should show a dramatic turnup at low count rates due to early evolutionary effects.

The total merging event rate predicted by our evolutionary model is $R \approx 2 \times 10^{-4}$ per year per $10^{11} M_\odot$, that is $\sim 10^7$ events/yr for the entire Universe, implying a factor of $\sim 10^3$ overproduction relative to the presently observed BATSE GRB rate of 0.8 events per day. This could be explained, for example, by a relativistic beaming in GRB sources (Paczyński 1994). The corresponding angle required to explain the GRB anisotropy that high is about $\delta\theta \approx 10^{-3}/\sqrt{R} \simeq 3°$ (Mao & Yi 1994). Taking this factor into account yields the expected total GRB rate $\sim 5000$ events per year for a limiting sensitivity lower by a factor of 3-10 than the presently exisitng BATSE limit. We note that use of another mean galactic density $n_G$ in the Universe would accordingly change the overal GRB rate $R \propto n_G$ in the Universe, but does not change the $\log N$–$\log S$ curve shape. Taking it less than 0.08 per cubic megaparsec would somewhat decrease the anisotropy required ($\propto R^{-1/2}$).

We conclude that the crucial test of the cosmological origin of GRBs would be observing the predicted increase of $\log N$–$\log S$ slope at smaller fluxes, inevitable due to early evolutionary effects. If the cosmological origin of GRB will be confirmed, the $\log N$–$\log S$ and $\langle V/V_{max} \rangle$ test could be used to independently estimating the cosmological parameters and tracing star formation history in galaxies.

## 5. Acknowledgements

The authors acknowledge Dr. A.G.Doroshkevich for useful discussions and the anonymous referee for comments. The work was partially supported by the INTAS grant 93-3364, the RFFR grant 94-02-04049a and by a grant of the Center for Cosmoparticle Physics COSMION. VML thanks the staff of Copenhagen University Observatory for its hospitality.

Fig. 1.— Temporal evolution of NS+NS (filled circles) and NS+BH (open circles) coalescence rates calculated for $2 \times 10^8$ binaries and normalized to a model elliptical galaxy with baryonic mass $10^{11} M_\odot$.

Fig. 2.— Confidence level contour lines (90%, 91%, ...) for $\omega^2$ test in the $\epsilon - \Omega(\Lambda)$ plane at $z_* = 2.25$ (a), in the $\epsilon - z_*$ plane for $\Omega(\Lambda) = 0.8$ (b) and in the $\Omega(\Lambda) - z_*$ plane for $\epsilon = 0.6$. A flat cosmologcal model and GRB spectral index $s = 1.5$ with the source evolution as in Fig. 1 is adopted.

Fig. 3.— $\log N$–$\log S$ diagram simulated for sources evolving as shown in Fig. 1 in a flat cosmological model with the best-fit vacuum energy $\Omega_\Lambda = 0.75$ and parameter $\epsilon = 0.5$.

Fig. 4.— Dependency of the differential $\langle V/V_{max} \rangle$-test on sample limit redshift $z$ for different epochs of the primordial star formation $z_*$

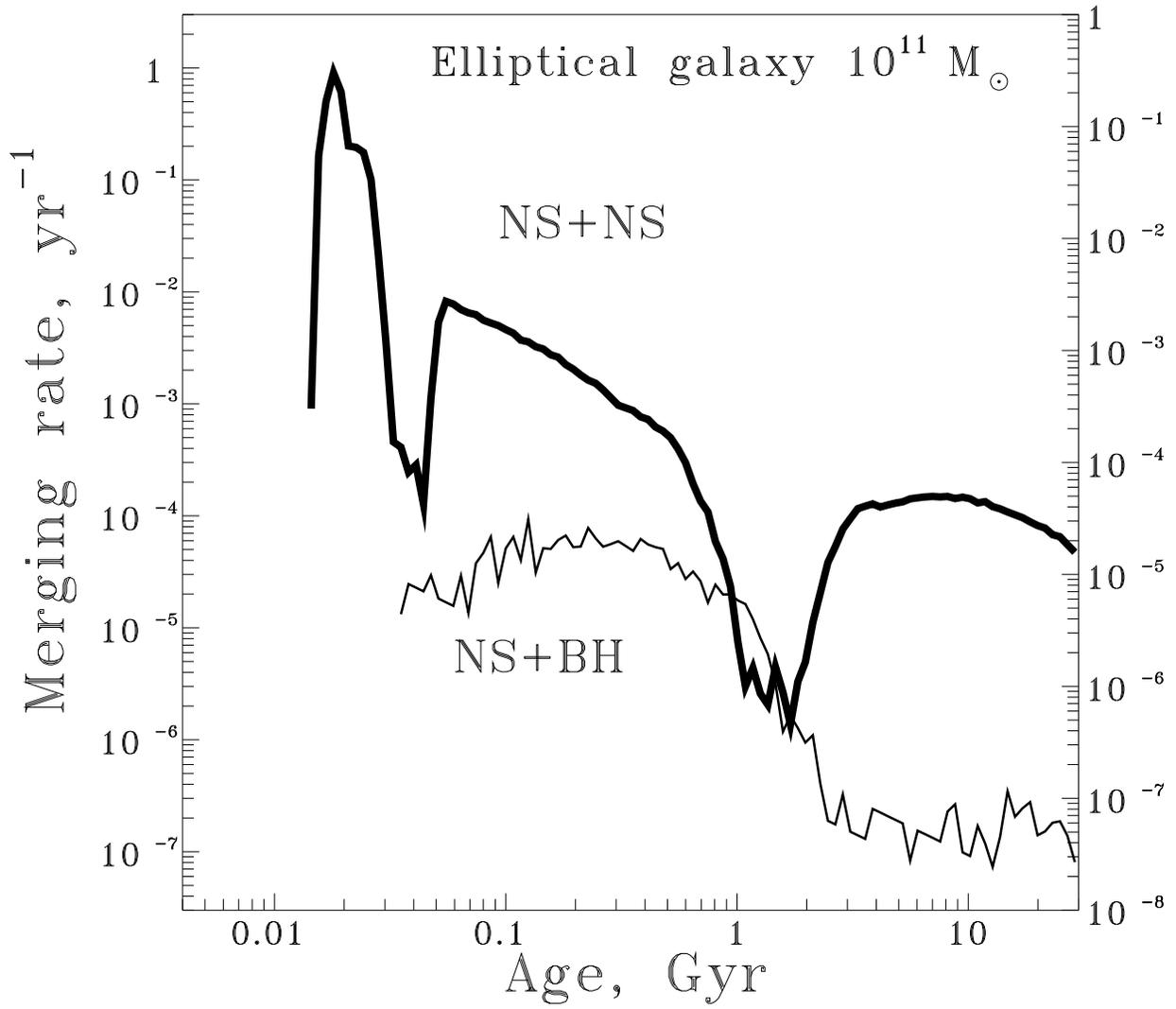

Figure 1.

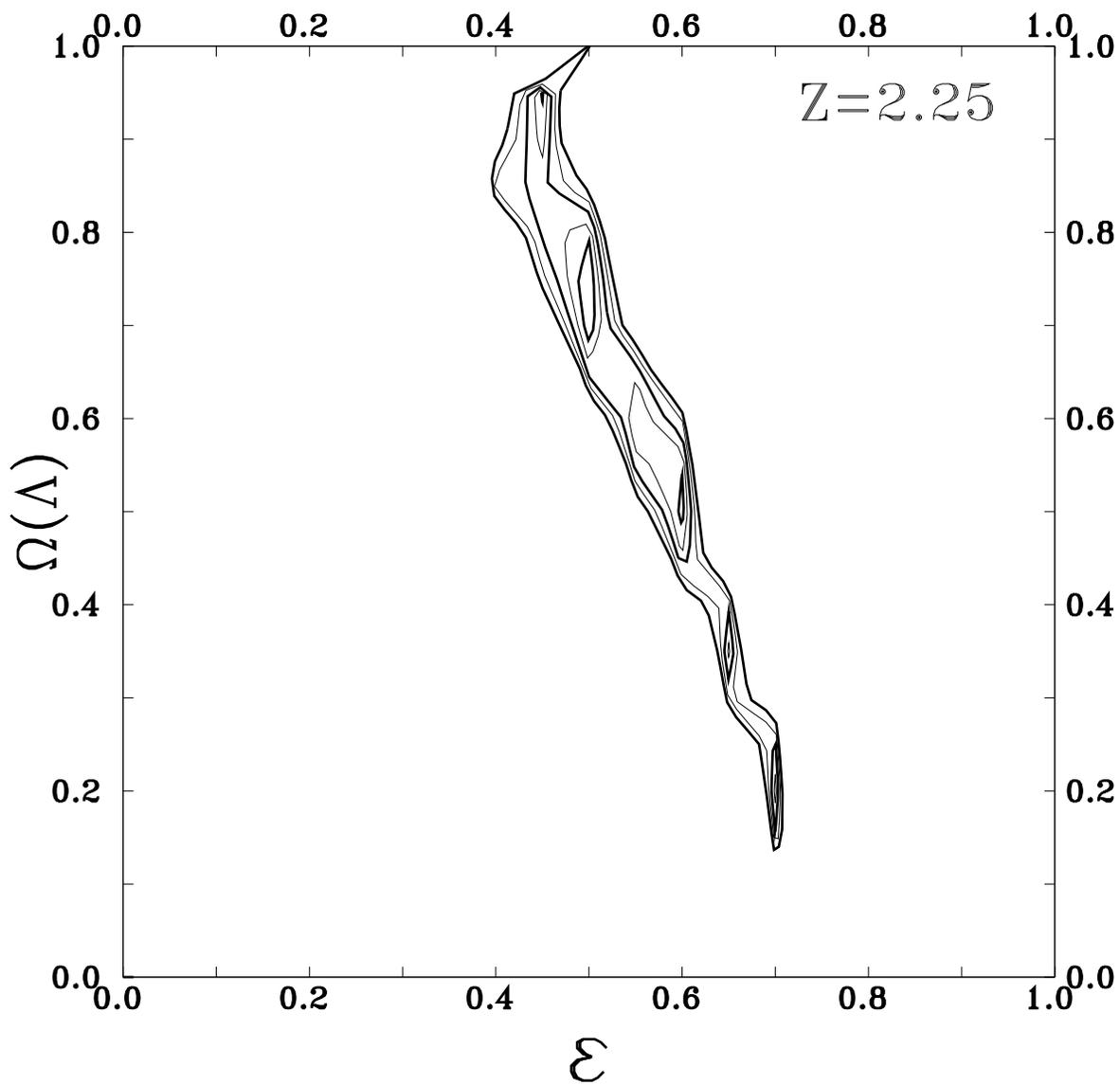

Figure 2a.

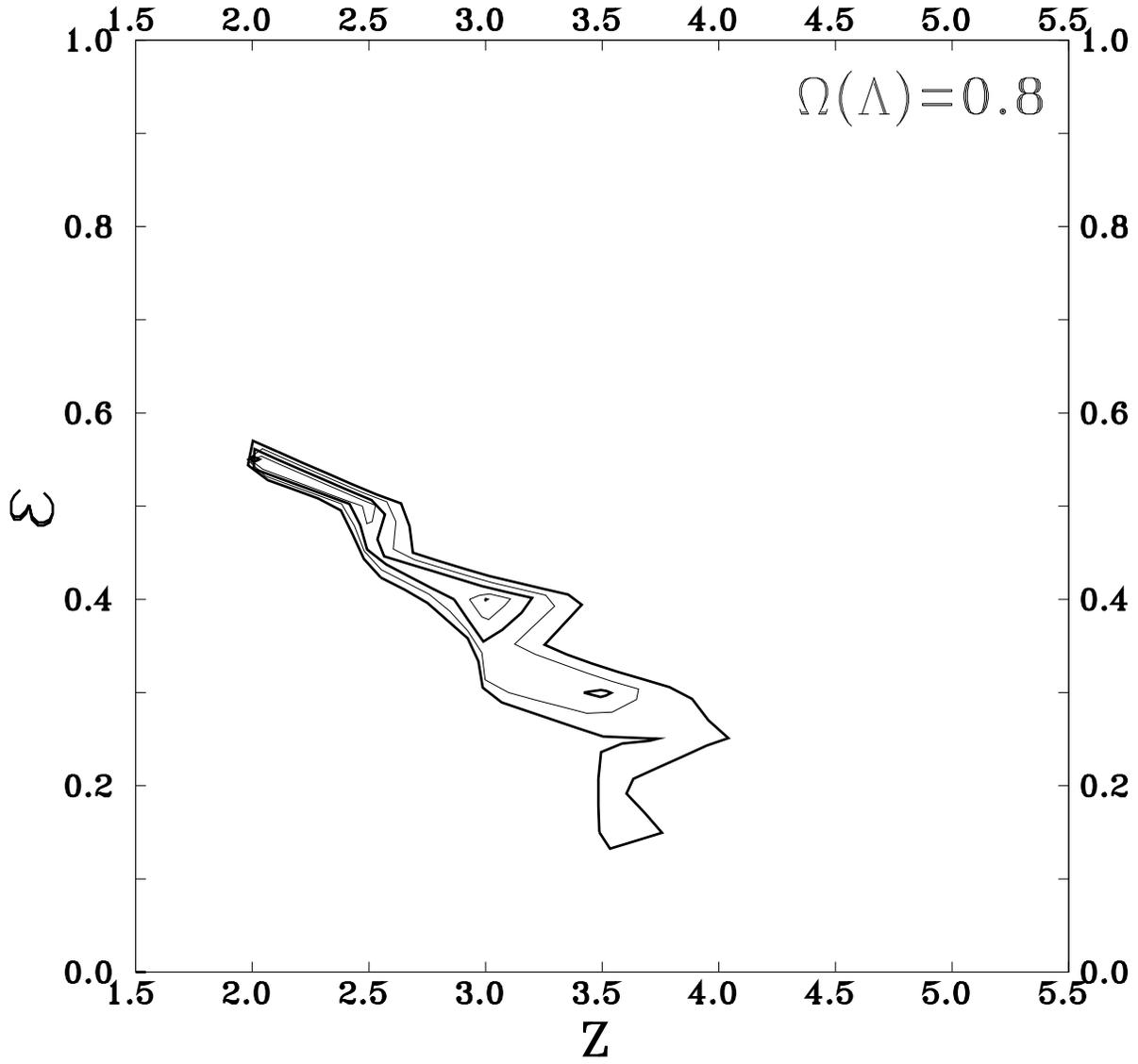

Figure 2b.

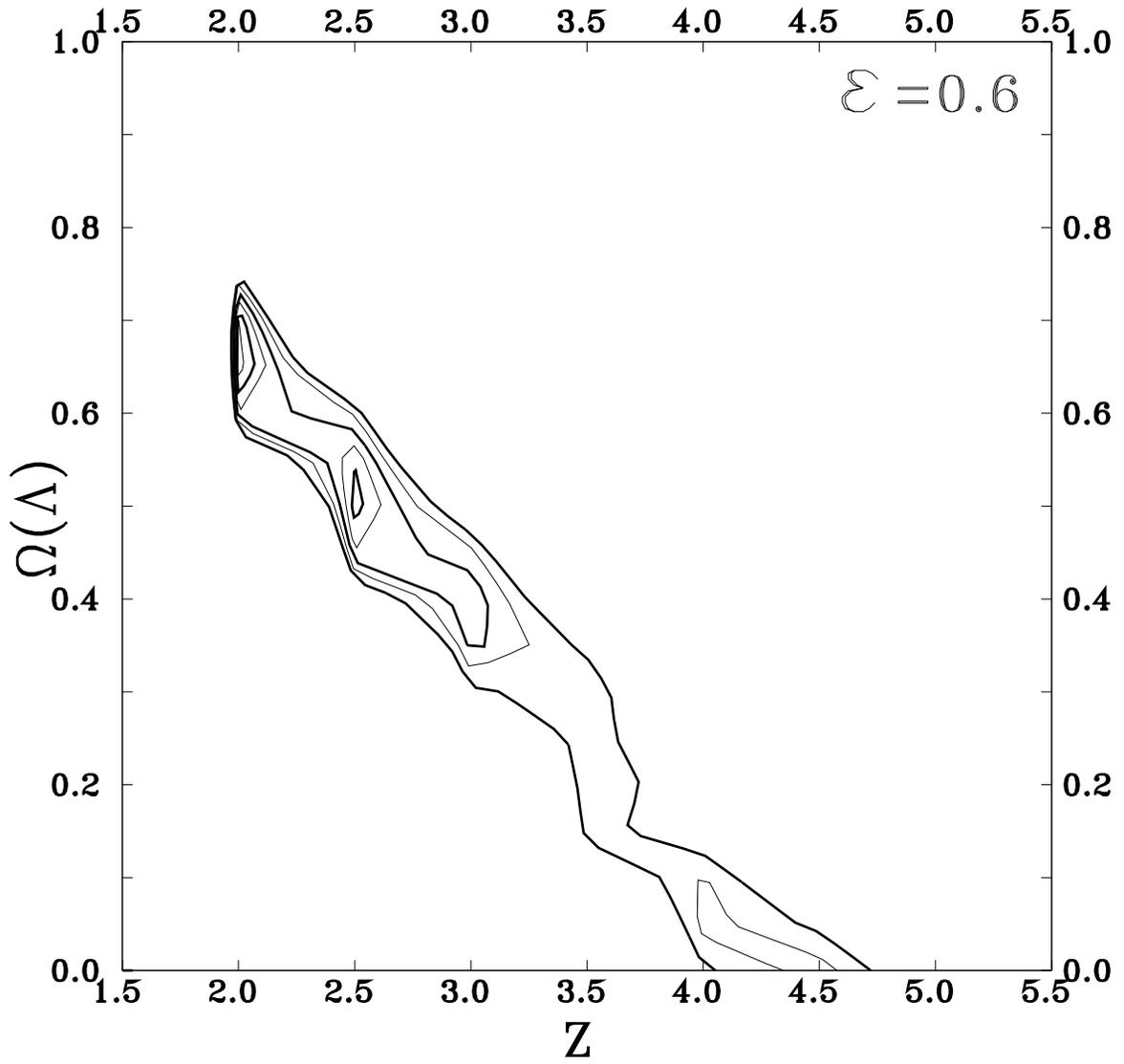

Figure 2c.

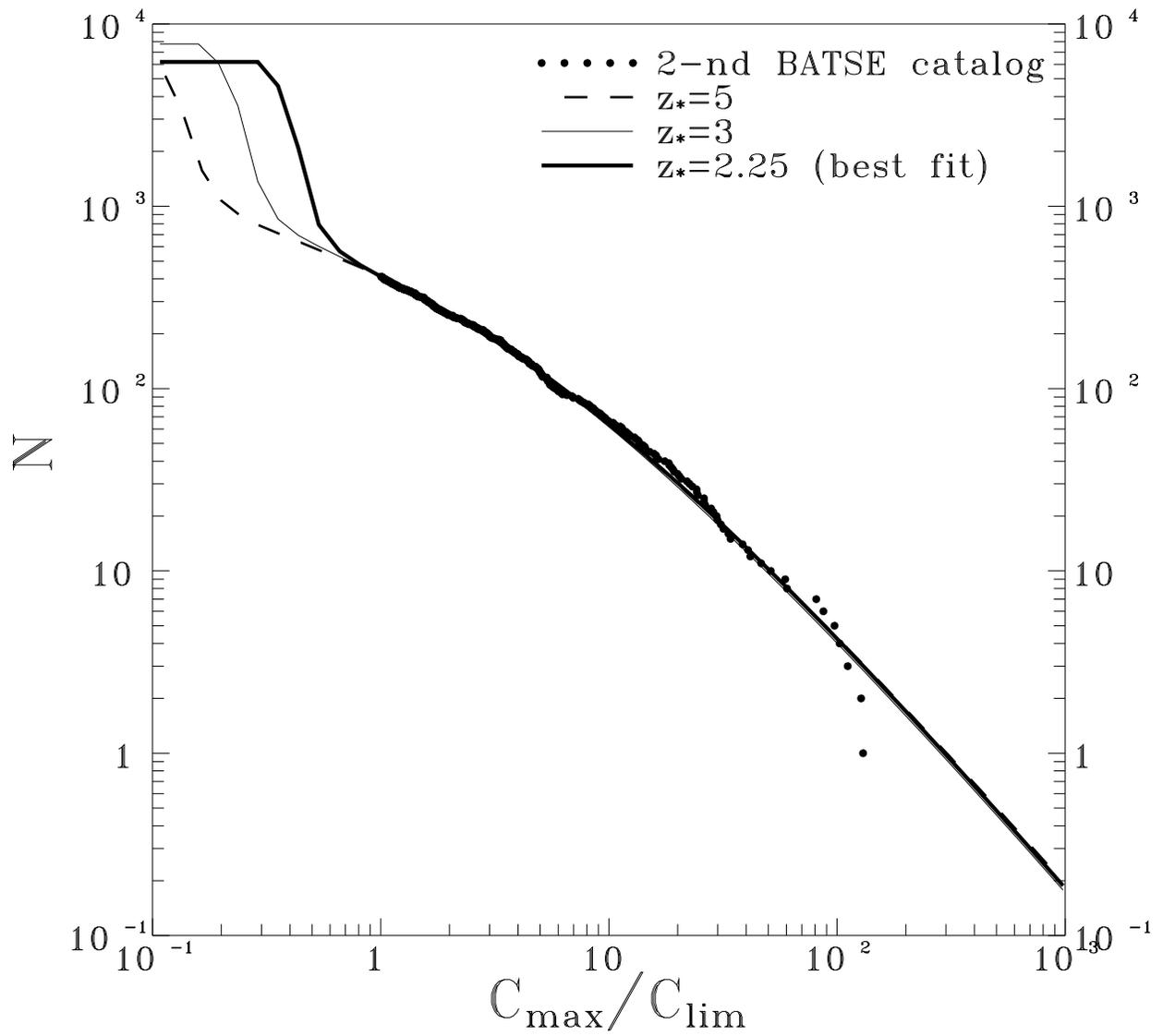

Figure 3.

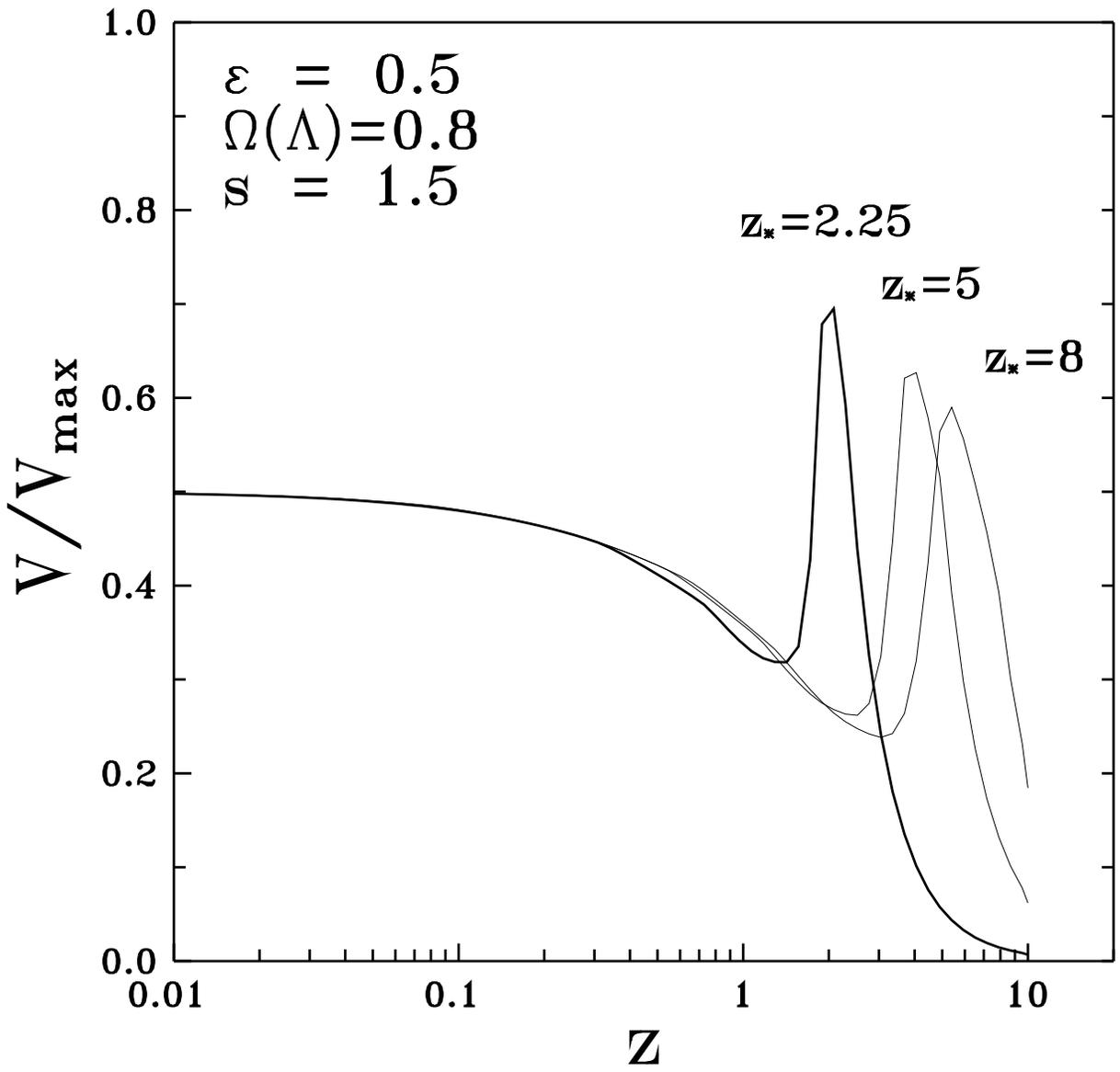

Figure 4.